\documentclass[12pt]{article}
\usepackage{amsmath}
\usepackage{color}
\usepackage{amssymb}
\usepackage{bm}
\usepackage{graphicx}

\usepackage{dcolumn}% Align table columns on decimal point
\usepackage{indentfirst}

\usepackage[T1]{fontenc} % if needed

\newcommand{\f}{\begin{equation}}
\newcommand{\ff}{\end{equation}}
\newcommand{\fa}{\begin{eqnarray}}
\newcommand{\ffa}{\end{eqnarray}}

\newcommand{\nn}{\nonumber}

\makeatletter\@addtoreset{equation}{section}
\makeatother
\allowdisplaybreaks[2]
\begin{document}
\begin{titlepage}

\begin{flushright}
\phantom{preprint no.}
\end{flushright}
\vspace{0.5cm}
\begin{center}
{\Large \bf
Scattering cross section of the long gravitino wave \\
\vspace{2mm}
in Schwarzschild spacetime
}
\lineskip .75em
\vskip0.5cm
{\large Yikang Xiao${}^{1}$ and Wenbin Lin${}^{2,\,3,\,*}$}
\vskip 2.5em
${}^{1}$ {\normalsize\it School of Nuclear Science and Technology, \\
University of South China, Hengyang, 421001, China\\}
\vskip 1.0em
${}^{2}$ {\normalsize\it School of Mathematics and Physics, \\
University of South China, Hengyang, 421001, China\\}
\vskip 1.0em
${}^{3}$ {\normalsize\it School of Physical Science and Technology, Southwest Jiaotong University, \\ Chengdu, 610031, China\\
}
\vskip 1.0em
${}^{*}$ {\normalsize\it Email: lwb@usc.edu.cn\\}
\vskip 1.0em
\vskip 3.0em
\end{center}
\begin{abstract}
%We investigate the scattering of gravitino wave by Schwarzschild field in long-wavelength limit. Using perturbative techniques, we compute the differential scattering cross section through the simulating wave function method. Our work demonstrates that the spin-dependence of the gravitino scattering cross section aligns with the behaviors for the scalar, neutrino, and electromagnetic waves.
%
%We investigate the scattering of gravitino waves by a Schwarzschild gravitational field in the long-wavelength limit. By employing perturbative techniques together with the simulating wave function method, we calculate the differential scattering cross section of the gravitino. Our analysis reveals that the spin dependence of the gravitino scattering cross section exhibits the same characteristic pattern as those of scalar, neutrino, and electromagnetic waves, concluding the long-wavelength scattering problem of the supersymmetric particles in the Spherically symmetry field. 
%
We investigate the scattering of gravitino wave in a Schwarzschild gravitational field. Employing the simulating wave function method within the framework of perturbative techniques, we derive the differential scattering cross section for the gravitino wave in the long-wavelength limit. It is proven that the cross section of gravitino wave follows the same spin-dependent pattern as those of scalar, neutrino, and electromagnetic waves. %This work provides a quantitative conclusion for the long-wavelength scattering problem of supersymmetric particles in a spherically symmetric gravitational field.  
\end{abstract}
\end{titlepage}

\section{Introduction}
\label{sec:intro}

Quantum waves encompass both matter and massless waves. According to the Coleman-Mandula theorem, massless waves can be classified by spin $s$: scalar waves ($s\!=\!0$), fermionic waves ($s\!=\!1/2$), electromagnetic waves ($s\!=\!1$), gravitino waves ($s\!=\!3/2$), and gravitational waves ($s\!=\!2$). The propagation of these waves in a gravitational field has long been a classical problem in physics.

Extensive research has focused on the gravitational scattering of massless waves with various spins by compact objects such as black holes, particularly emphasizing the differential scattering cross section~\cite{anchez:1976qua}. Typically, the incident wave is assumed to be a weak plane wave. The scattering process is characterized by the dimensionless parameter:
%The quantum waves include the matter  dimensionless parameter,
\begin{equation}\label{Mo}
  M \omega = \pi r_S /\lambda~,
\end{equation}
using natural units ($G\!=\!c\!=\!1$), where
$M$ and $r_S$ is the mass and Schwarzschild radius of the compact body, $\omega$ and $\lambda$ are the angular frequency and wavelength of the incident wave respectively.$\omega=\omega_{\text(in)}$

In the long-wavelength regime ($M \omega \!\ll\! 1$), the differential cross section in a Schwarzschild field is given by
%where the natural unit ($G=c=1$) is  Schwarzschild field as follows
~\cite{Westervelt:1971wq,Peters:1976vwg,anchez:1976qua,Matzner:1977bg,Logi:1977,Doran:2002xmb,Dolan:2019eyr,Dolan:2007kr}
\begin{equation}\label{dsdo}
\lim_{M\omega\rightarrow 0}  \Big( \frac{1}{M^2}\frac{d\sigma}{d\Omega} \Big) = \left\{
\begin{array}{ll}
\frac{1}{\sin^4(\theta/2)}~, ~~& \text{for} ~~s=0~,\\
\frac{\cos^2(\theta/2)}{\sin^{4}(\theta/2)}~,~~ &   \text{for}~~ s=\frac{1}{2}~, \\
\frac{\cos^4(\theta/2)}{\sin^{4}(\theta/2)}~,~~ &  \text{for}~~  s=1~,  \\
%\frac{\cos^4(\theta/2)}{\sin^{4}(\theta/2) ~~& s=1 , \\
\frac{\cos^8(\theta/2)}{\sin^{4}(\theta/2)}+\sin^4(\theta/2)~, ~~ &  \text{for} ~~s=2~, %
\end{array} \right.
\end{equation}
Here, $\sigma$ is the scattering cross section and $\Omega$ is the solid angle associated with the scattering angle $\theta$.

For the case of $s\!=\!\frac{3}{2}$, although the equations-of-motion for gravitino in various gravitational fields have been explored~\cite{Castillo:1990pw,Silva-Ortigoza1992}, the corresponding cross section remains uncalculated. This omission is likely due to the complexity of spinor algebra involved in the fermionic spin-$\frac{3}{2}$ case, which is considerably more complex than for lower spin values.

Scattering calculations are commonly performed using perturbative techniques such as Feynman diagrams. For example, Guadagnini employed Feynman diagrams to compute low-energy graviton scattering in the background of classical massive bodies at first post-Minkowskian (1PM) order~\cite{Guadagnini:2008vwg}. Higher-order gravitational scattering effects for scalar particles have also been studied~\cite{Festuccia:2018xjb}. Further studies have addressed wave scattering with various spins except spin-$\frac{3}{2}$ in different gravitational fields~\cite{Vines:2021wq,Sen:2018pw,Cheung2018}.
%the scattering amplitude caused by soft photons and soft gravitons~\cite{Sen:2018pw}, as well as the scattering problem in the gravitational field with the method of classical potential~\cite{Cheung2018}.
 Boundary-to-bound state transitions have been investigated via holographic methods~\cite{Porto:2019xjb}.
If we employ certain mathematical physics methods, it might make the calculations appear somewhat simpler. The methods we have considered include the partial wave method in spherical coordinates and Feynman diagrams. These methods have their unique advantages in handling complex problems. The mathematical rigor of the partial wave method in spherical coordinates can be guaranteed. However, this partial wave method requires the Teukolsky equation with $s =-3/2$. Understanding the Teukolsky equation demands considerable mathematical background, such as the Newman-Penrose equations, which poses difficulties for some readers. Furthermore, $\Gamma$ and $\Psi$ functions appear in the calculation process of the partial wave method in spherical coordinates, and these functions have little connection with our first-order perturbation theory. Therefore, after incorporating the details of the mathematical background, the partial wave method in spherical coordinates does not become simpler.
If we use Feynman diagrams, we face the problem of why following the Feynman rules yields the scattering amplitude. The physical problem in this paper is not as easily comparable with experiments as quantum electrodynamics. Explaining the Feynman rules would involve issues such as ghost fields and Becehi-Rouet-Store-Tyutin quantization, which are unnecessary complications. After resolving these complications, we would still encounter the problem of strictness in supergravity. To avoid this series of difficulties, we consider adopting first-order perturbation calculations directly in Cartesian coordinates.

In this work, we compute the long-wavelength scattering cross section of the gravitino in Schwarzschild spacetime. Using a Cartesian coordinate system as in the work of Doran and Lasenby~\cite{Doran:2002xmb}, we apply first-order classical perturbation theory—equivalent to Feynman diagram calculations—and include an analysis of the spin $\pm \frac{1}{2}$ degrees of freedom. Our approach bypasses the need for explicit Feynman diagrams by directly utilizing the method of wave function simulation, which is similar to Archimedes' method for calculating buoyancy.

\section{Basis for spinor algebra}

The action on gravitino in $N=1$ super-gravity is similar to the Weyl Fermion action. Gravitino satisfies the Rarita-Schwinger equation~\cite{Chandrasekhar:2018xjb}
\f
\Gamma^{[\mu B'}_C \Gamma^{\nu C}_{D'} \Gamma^{\alpha]D'}_A  \nabla_\mu \psi_\alpha^A=0~,\label{RS}
\ff
with
\f
\Gamma^{\mu B'}_C=e_{(\tau)}^\mu \Gamma^{(\tau) B'}_C~,
\ff
where $\Gamma^{(\tau) B'}_C$ is the $2\times2$ part of Dirac matrix. $e_{(\tau)}^{\mu}$ is tetrad basis. The prime denote the complex conjugate. In the superscripts and subscripts, the capital letters denotes the spinor indices, Greek letters, e.g., $\mu,~\nu,~\alpha$, denote the 4-dimensional spacetime indices, Greek letters with parentheses, e.g., $(\tau),~(\beta),~(\gamma)$, denote the 4-dimensional tetrad indices, Latin letters, e.g., $i,~j,~k$, denote the 3-dimensional space indices, and the Latin letters with parentheses, e.g., $(i),~(j),~(k),$, denote the 3-dimensional indices for the tetrad. The gravitino wave function is represented by 
$\psi_\alpha^A\!=\!e^{(\tau)}_{\alpha} \psi_{(\tau)}^A$ with $A$ being the spinor index ($\downarrow$ or $\uparrow$). The notation ``$[\cdot\cdot\cdot]$'' denotes the total anti-symmetrization of the Greek-letter indices.

Rarita-Schwinger equation is equivalent to
\fa
\sigma_{AB'(\tau)} e^\mu_{[(\beta)}\nabla_\mu \psi^A_{(\gamma)]} + \sigma_{AB'(\beta) } e^\mu_{[(\gamma)}\nabla_\mu \psi^A_{(\tau)]}+\sigma_{AB'(\gamma)}e^\mu_{[(\tau)}\nabla_\mu \psi^A_{(\beta)]}=0~, \label{RS1}
\ffa
where $\sigma_{AB'(\alpha) }$ is the Pauli matrix
\begin{equation}
 \sigma_{AB'(0) } = \begin{pmatrix}
    1 & 0 \\
    0 & 1
  \end{pmatrix}~,~~
  \sigma_{AB'(1) } = \begin{pmatrix}
    0 & 1 \\
    1 & 0
  \end{pmatrix}~,~~
  \sigma_{AB'(2) } = \begin{pmatrix}
    0 & -\text{i} \\
    \text{i} & 0
  \end{pmatrix}~,~~
\sigma_{AB'(3) } = \begin{pmatrix}
    1 & 0 \\
    0 & -1
  \end{pmatrix}~,
\end{equation}
where $\text{i}$ denotes the imaginary unit.

\subsection{Spinor field in Minkowskian spacetime}
We first solve the Rarita-Schwinger equation for a flat background. Because Fermion generation and annihilation are not considered, the Grassmann number is not used. We assume that the incident wave is along the $z$-direction.
Let $\psi_\alpha^A=e^{(\tau)}_{\alpha}\Phi_{(\tau)}^{A} exp\{\text{i} \omega (z-t)\}$, with $\Phi_{(\tau)}^{A}$ being independent of spatial location. $e^{(\tau)}_{\alpha}$ is the tetrad of spacetime. There are three excitation modes
\fa
&& \hskip -0.5cm \text{Mode 1:}~\Phi_{(1)}^{\downarrow }=-i\Phi_{(2)}^{\downarrow }\neq0,~\Phi_{(0)}^{\uparrow }=\Phi_{(0)}^{\downarrow }=\Phi_{(1)}^{\uparrow }=\Phi_{(2)}^{\uparrow }=\Phi_{(3)}^{\uparrow }=\Phi_{(3)}^{\downarrow }=0~,\\
&& \hskip -0.5cm \text{Mode 2:}~\Phi_{(0)}^{\uparrow }=-\Phi_{(3)}^{\uparrow }\neq0,~\Phi_{(0)}^{\downarrow }=\Phi_{(1)}^{\downarrow }=\Phi_{(1)}^{\uparrow }=\Phi_{(2)}^{\downarrow }=-i\Phi_{(2)}^{\uparrow }=\Phi_{(3)}^{\downarrow }=0~,\\
&& \hskip -0.5cm \text{Mode 3:}~\Phi_{(0)}^{\downarrow }=-\Phi_{(3)}^{\downarrow }\neq0,~\Phi_{(0)}^{\uparrow }=\Phi_{(1)}^{\downarrow }=\Phi_{(1)}^{\uparrow }=-i\Phi_{(2)}^{\uparrow }=\Phi_{(2)}^{\downarrow }=\Phi_{(3)}^{\uparrow }=0~,
\ffa %$\uparrow$ is 1 and $\downarrow$ is 2.
It is well-known that Mode 1 is a physical excitation, while Mode 2 and 3 are caused by the local supersymmetry transformation thus are not.

From now on, we only discuss Mode 1, in which $\Phi_{(\tau)}^{A}$ can be written in the form of direct product
\begin{equation}
\Phi_{(\tau)}^{A}=\varepsilon  m_{(\tau)} \xi_{(\text{in})}^A~,\label{z4}
\end{equation}
where $\varepsilon$ is a real number, and is set as 1 in the linear case. $\xi_{(\text{in})}^\uparrow=0$ and $\xi_{(\text{in})}^\downarrow=1$. $m_{(\tau)} $ is similar to Newman-Penrose basis $\bar{m}$,
\fa
&& m_{(0)}=0~,\label{m_0}\\
&& m_{(i)}=(1/\sqrt2) \sigma_{AB'(i)} \sigma_{CD'}^{(0)} \epsilon^{D'B'} \xi_{(\text{in})}^A \xi_{(\text{in})}^C~,\label{m_i}
\ffa
\color{magenta}
%Latin letter $i,j,k,l,m,n,o,p,q$ denote the 3-dimensional space indices.
\color{black}
After direct calculation, we have $m _{(1)}=1/\sqrt2,~m _{(2)}=i/\sqrt2,~m _{(3)}=0$.

The incident wave vector $k_{(\text{in})\alpha}$ satisfies
\fa
&& k_{(\text{in})\alpha}e_{(\tau)}^\alpha \sigma_{AB'}^{(\tau)} \xi_{(\text{in})}^A=0~,  \label{z5} \\
&& m_{(\tau)}e_\alpha^{(\tau)} k_{(\text{in})}^\alpha=m_{(\tau)} m^{(\tau)}=0~.\label{z6}   \\
&& \sigma_{CD'}^\beta k_{(\text{in})\beta} \epsilon^{D'B'} \sigma_{AB'}^\alpha m_\alpha=2\sqrt{2} \omega\epsilon_{CD} \epsilon_{AB} \xi_{(\text{in})}^D \xi_{(\text{in})}^B~,   \label{z7}
\ffa
where $\epsilon^{D'B'}$ is an invariant tensor
\fa
&& \epsilon^{\uparrow'\downarrow'}=-\epsilon^{\downarrow'\uparrow'}=1~,\\
&& \text{elsewise,}~~ 0~.
\ffa

\subsection{Spinor field in spherically symmetric spacetime}
Here we only consider the first-order post-Minkowskian (1PM) solution for Rarita-Schwinger equation in Schwarzschild spacetime. The Schwarzschild metric in the 1PM approximation can be written as
\f
%ds^2=-(1-2M/r ) dt^2+(1-2M/r )^{-1} dr^2+r^2 d\Omega^2
ds^2=-\Big(1-2\frac{M}{r}\Big) dt^2+ d\bm{x}\cdot d\bm{x} +\frac{2M}{r^3} (\bm{x} \cdot d\bm{x})^2 ~,
\ff
where $M$ denotes the mass of Schwarzschild gravitational source, and $r=|\bm{x}|$. The natural units in which $G\!=\!c\!=\!1$ are adopted.
%At least, for low-density objects, such an expansion is valid. Later, we will discuss the case of black holes.

We choose tetrads $e_\mu^{(\tau) } dx^\mu$ everywhere so that $e_\mu^{(\tau) } e_\nu^{(\beta) } \eta_{(\tau)(\beta)} =g_{\mu\nu}$,
\begin{eqnarray}
&& e_0^{(0) }=1-M/r~,\\
&& e_0^{(i) }=e_i^{(0) }=0~,\\
&&  e_j^{(i) }=\delta_j^i+(M/r^3 ) x^i x_j~,\\
&& e_{(j)}^i=\delta_j^i-(M/r^3 ) x^i x_j~,
\end{eqnarray}
Taking the time and space derivatives of $e_\mu^{(\tau)}$, we have
\fa
&& e_{0,0}^{(0) }=e_{0,0}^{(i) }=e_{i,0}^{(0) }=e_{j,0}^{(i) }=e_{0,j}^{(i) }=e_{i,j}^{(0) }=0~,\\
&& e_{0,i}^{(0) }=(M/r^3 ) x_i~,\\
&& e_{j,k}^{(i) }=-3(M/r^5 ) x^i x_j x_k+(M/r^3 ) \delta_k^i x_j+(M/r^3 ) x^i \delta_{jk}~,
\ffa
where the subscript ``$,$'' denotes the derivative. $\delta$ is Kronecker sign.

In order to calculate the covariant derivative of wave function, we introduce a tensor $\lambda_{(\tau)(\beta)(\gamma)}$ with three tetrad indices as follows
\fa
&& \lambda_{(0)(0)(0)} =\lambda_{(i)(0)(j)} =\lambda_{(0)(i)(0)} =\lambda_{(0)(j)(i)} =0~,\\
&& \lambda_{(0)(0)(i)} =-e_{0,j}^{(0) }e_{(0)}^0 e_{(i)}^j=-(M/r^3 ) x_i~,\\
&& \lambda_{(k)(j)(i)} =e_{l,m}^{(j) } e_{(k)}^l e_{(i)}^m-e_{m,l}^{(j) } e_{(k)}^l e_{(i)}^m=(M/r^3 )(\delta_k^j x_i-\delta_i^j x_k )~,
\ffa

Ricci rotation coefficient $\gamma_{(\tau)(\beta)(\gamma)}$ is related to $\lambda_{(\tau)(\beta)(\gamma)}$ as
\f
 \gamma_{(\tau)(\beta)(\gamma)} =(\frac{1}{2}) (\lambda_{(\tau)(\beta)(\gamma)} +\lambda_{(\gamma)(\tau)(\beta)} -\lambda_{(\beta)(\gamma)(\tau)} )~,
\ff
i.e,
\fa
&& \gamma_{(0)(j)(0)} =\lambda_{(0)(0)(j)} =-(M/r^3 ) x_j~, \label{gamma1}\\
&& \gamma_{(k)(j)(i)}=(M/r )(\delta_i^k x_j-\delta_i^j x_k )~.\label{gamma2}
\ffa

The Ricci rotation coefficient also satisfies
\f
e_k^{(\gamma) } \gamma_{(\gamma)(\tau)(\beta)}  e_i^{(\beta) }=e_{(\tau)k;i}~,
\ff
where the subscript ``$;$'' denotes the covariant derivative.

At the same time, we also need two spin connections $\Upsilon_{A(\tau)}^B$ and $\Upsilon_{A'(\tau)}^{B'}$, which satisfy
\fa
&& (\Upsilon_{A(\tau)}^B )^*=\Upsilon_{A'(\tau)}^{B'}~,\\
&& \sigma_{CB'}^{ (\beta) }\Upsilon_{A(\tau)}^C + \sigma_{AC'}^{ (\beta) } \Upsilon_{B'(\tau)}^{C'} - \sigma_{AB'}^{ (\gamma) } \gamma_{(\gamma)(\tau)}^{(\beta) }=0~,
\ffa
where the superscript ``$*$'' denotes the complex conjugate. From these two equations we have
\f
\sigma_{(\delta)}^{AB'}(\sigma_{CB'}^{ (\beta) }\Upsilon_{A(\tau)}^C + \sigma_{AC'}^{ (\beta) } \Upsilon_{B'(\tau)}^{C'}- \sigma_{AB'}^{ (\gamma) } \gamma_{ (\gamma)(\tau)}^{(\beta) } )=0~.
\ff

We can choose
\f
\Upsilon_{A(\tau)}^B=\frac{1}{4} \gamma_{(\beta)(\delta)(\tau)}  \sigma_{AD'}^{(\beta) } \sigma^{ BD'(\delta) }~,
\ff
then we have
\fa
&& \Upsilon_{A'(\tau)}^{B'}=\frac{1}{4} \gamma_{(\beta)(\delta)(\tau)}  \sigma_{DA'}^{ (\beta)}  \sigma^{ DB'(\delta) }~,\\
&& \sigma_{(\delta)}^{AB'} \sigma_{AB'}^{(\beta) }=-2\delta_{d}^{b}~,\\
&& \gamma_{(\delta)(\tau)}^{(\beta) }=-\frac{1}{2}\sigma_{(\delta)}^{AB'} (\Upsilon_{A(\tau)}^C \sigma_{CB'}^{ (\beta) }+\sigma_{AC'}^{ (\beta) } \Upsilon_{B'(\tau)}^{C'} )~.
\ffa

We employ the perturbation method to calculate the scattering problem. The 1PM solution is the summation of the zero-order solution (the incident field) and the first-order correction
\f
\psi_\alpha^{A}=e_\alpha^{(\tau) } (\psi_{\{0\}(\tau)}^{A}+\psi_{\{1\}(\tau)}^{A})~,\label{psi}
\ff
where the notation ``$\{0\}$'' represents the flat spacetime, and ``$\{1\}$'' represents the first-order perturbation on the flat spacetime. For the zero-order solution, we can write down
\f
\psi_{\{0\}(\tau) }^{A} = \Phi_{(\tau)}^{A} \exp\{\text{i} \omega (z-t)\}~.
\ff

Take a distant past $t \!=\!-T/2$ as the boundary, starting the evolution from the initial value $\psi_\alpha^{A}(-T/2)$. 
Take $\psi_\alpha^{A}(-T/2)\approx\delta_{\alpha}^{(\tau)} \psi_{\{0\}(\tau) }^{A}(-T/2)$
They are not equivalent because of the constraint equation. As will be shown later, this difference has a negligible impact on the scattering amplitude due to both group representation and coherence considerations.

The first-order correction is key contribution of this work. Substituting Eq.\,(\ref{psi}) into Eq.\,(\ref{RS1}), we have
\begin{eqnarray}
&&   \sigma_{BA'[(\tau)}^{~}   \partial_\beta \psi_{\{1\}(\gamma)]}^B + \sigma_{BA'[(\gamma)}^{~}  \partial_\tau \psi_{\{1\}(\beta)]}^B + \sigma_{BA'[(\beta)}^{~}   \partial_\gamma \psi_{\{1\}(\tau)]}^B \nn\\
&& ~~~~ + \,3   (Z_1)_{(\tau)(\beta)(\gamma)A'} + 3 (I_1)_{(\tau)(\beta)(\gamma)A'} =0~,\label{psiEoB}
\end{eqnarray}
where
\fa
&& (Z_1)_{(\tau)(\beta)(\gamma)A'} =  \gamma_{[(\tau)(\beta)}^{(\delta) } \sigma_{BA'(\gamma)]} \psi_{\{0\}(\delta)}^B - \sigma_{BA'[(\gamma) } \Upsilon_{C(\beta)}^B \psi_{\{0\}(\tau)]}^C~,\label{I40}\\
&& (I_1)_{(\tau)(\beta)(\gamma)A'} = -\Delta e_{[(\beta)}^\mu \sigma_{BA'(\gamma) } \partial_\mu \psi_{\{0\}(\tau)]}^B~,\label{I6}
\ffa
with $\Delta e_{(\beta)}^\mu\equiv e_{(\beta)}^\mu-\delta_{b}^\mu$~.

%$(Z_i)$ with $i=1,2,\cdot\cdot\cdot$

Now, the calculation of the differential scattering cross-section can be transferred to analyzing the contribution of $(Z_1)_{(\tau)(\beta)(\gamma)A'}$ and $(I_1)_{(\tau)(\beta)(\gamma)A'}$.  More specifically, $(Z_1)_{(\tau)(\beta)(\gamma)A'}$ is induced by Ricci rotation coefficient and spin connection.  $(I_1)_{(\tau)(\beta)(\gamma)A'}$ is induced by the perturbed tetrad basis.

\subsubsection{The effects of Ricci rotation coefficient and spin connection}
First, we consider the contribution of $(Z_1)_{(\tau)(\beta)(\gamma)A'}$.  It can be seen from the RHS of Eq.\,(\ref{I40}) that $(Z_1)_{(\tau)(\beta)(\gamma)A'}$ has the full anti-symmetrization for the indices $(\tau),(\beta),(\gamma)$, we arrive at the conclusion that  $(Z_1)_{(i)(j)(k)A'}$ belongs to the fundamental spinor representation of SO(3) group, thus it cannot make contribution to the scattering of the massless wave with spin $\frac{3}{2}$.

For $(Z_1)_{(i)(j)(0)A'}$, we substitute Eqs.\, (\ref{z7}), (\ref{gamma1}) and (\ref{gamma2}) into Eq.\,(\ref{I40}), we can obtain
\begin{eqnarray}
&&(Z_1)_{(i)(j)(0)A'}=\frac{5iM}{24\sqrt2 \pi^2}\sigma_{MD'[(i) } \sigma_{NO'(j)]}\epsilon^{D'K'} \epsilon^{O'S'} \epsilon^{MN} \nonumber\\
 &&~~~~\times\, \sigma_{GS'}^{ (0) } \sigma_{EK'}^{(l)} \sigma_{BA'}^{ (0) } \xi_{(\text{in})}^E  \xi_{(\text{in})}^G \xi_{(\text{in})}^B\int \frac{k_{\text{(tran)}l}}{\bm{k_{\text{(tran)}}}^2 } \text{e}^{\text{i}(k_{\text{(tran)}\alpha}+k_{(\text{in})\alpha} ) x^\alpha }d^3k_{\text{(tran)}}\nonumber\\
&&~~~~+\,2\sigma_{MD'[(i) } \sigma_{NO'(j)]}  \epsilon^{D'K'} \epsilon^{O'S'} \epsilon^{MN} \big[\epsilon_{K'A'} (B_1)_{S'}+\epsilon_{S'A'} (B_2)_{K'}\big]~,\label{(Z_1)_{(i)(j)(0)A'}}
\end{eqnarray}
where $k_{\text(in)\alpha}$ is the ingoing wave vector with $k_{\text(in)0}=-\omega_{(\text{in})}$, and $k_{\text{(tran)}\alpha}$ is the internal wave vector in Feynman diagram with $k_0=0$. The variables
$(B_1)_{S'}$ and $(B_2)_{K'}$ denote the complex spinors with spin $\frac{1}{2}$. Since they cannot contribute to the scattering of the massless wave with spin $\frac{3}{2}$, so we do not bother to give their expressions. The integral exhibits ultraviolet divergence, but this divergence can be regularized using a low-pass filter.

Let
\begin{eqnarray}
&&\tau_{ijA'}=2\sigma_{MD'[(i) } \sigma_{NO'(j)]}\epsilon^{D'K'}\!\epsilon^{O'S'} \!\epsilon^{MN} \! J_{S'K'A'}\text{e}^{\text{i}(k_{\text{(tran)}\alpha}+k_{(\text{in})\alpha} ) x^\alpha }~,\label{a}
\end{eqnarray}
with
\begin{eqnarray}
J_{S'K'A'}\equiv \frac{5iM k_{\text{(tran)}l} \xi_{(\text{in})}^E \xi_{(\text{in})}^G \xi_{(\text{in})}^B}{144\sqrt2 \pi^2{\bm{k}^2_{\text{(tran)}}} }(\sigma_{GS'}^{(0)} \sigma_{EK'}^{(l)} \sigma_{BA'}^{ (0) }  \! + \!\sigma_{GK'}^{(0)} \sigma_{EA'}^{(l)} \sigma_{BS'}^{ (0) }   \!+\!\sigma_{GA'}^{(0)} \sigma_{ES'}^{(l)} \sigma_{BK'}^{ (0) }  ),\label{J}
\end{eqnarray}
%then
%\begin{eqnarray}
%&& \tau_{ijA'}\equiv \frac{5iM}{72\sqrt2 \pi^2}\sigma_{MD'[(i) } \sigma_{NO'(j)]}\epsilon^{D'K'} \epsilon^{O'S'} \epsilon^{MN}\frac{\xi_{(\text{in})}^E \xi_{(\text{in})}^G \xi_{(\text{in})}^B  k_{l}}{k^{\beta} k_{\beta}} \nonumber\\

then
\begin{eqnarray}
&& (Z_1)_{(i)(j)(0)A'}=\int \tau_{ijA'} d^3k_{\text{(tran)}} \nn\\
&& ~~~~ +\, 2 \sigma_{MD'[(i) } \sigma_{NO'(j)]} \epsilon^{D'K'} \epsilon^{O'S'} \epsilon^{MN} \big[\epsilon_{K'A'} (B_3)_{S'}\!+\!\epsilon_{S'A'} (B_4)_{K'}\big]~.\label{(}
\end{eqnarray}
Similarly, $(B_3)_{S'}$ and $(B_4)_{K'}$ denote the complex spinors with spin $\frac{1}{2}$, which cannot contribute to the scattering of the massless wave with spin $\frac{3}{2}$, so we do not bother to give their expressions (in the below, we will not mention this again for $(B_n)_{S'}$ and $(B_n)_{K'}$ with arbitrary $n$ due to the same reason). %for $(B_i)_{S'}$ and $(B_i)_{K'}$ with arbitrary $i$).

For the direction unit vector $(\sin\theta, 0, \cos\theta)$, we have ${\bm{k}^2_{\text{(tran)}}=4\omega_{(\text{in})}^2 \sin^2 (\theta/2)}$, and
\begin{eqnarray}
&& \hskip -1.25cm  J_{S'K'A'}=\frac{5\text{i}M k_{\text{(tran)}l} \xi_{(\text{in})}^E\xi_{(\text{in})}^G \xi_{(\text{in})}^B  \big(\sigma_{GS'}^{ (0) } \sigma_{EK'}^{(l)} \sigma_{BA'}^{ (0) }\!+\!\sigma_{GK'}^{ (0) } \sigma_{EA'}^{(l)}  \sigma_{BS'}^{ (0) } \!+\!\sigma_{GA'}^{(0) } \sigma_{ES'}^{(l)} \sigma_{BK'}^{ (0) } \big) }{576\sqrt2 \pi^2 \omega^2 \sin^2 (\theta/2) }.\label{hskip}
\end{eqnarray}

Finally, we can formulate the contribution of $(Z_1)_{(i)(j)(0)A'}$ to the massless wave with spin $\frac{3}{2}$ as
%can be written as
\f
2 \sigma_{MD'[(i) } \sigma_{NO'(j)]}\epsilon^{D'K'}\!\epsilon^{O'S'} \!\epsilon^{MN} J_{S'K'A'} ~.\label{2}
\ff
%\begin{eqnarray}
%\textcolor[rgb]{1.00,0.00,0.00}{ (Z_1)_{(i)(j)(0)A'} = 2 \sigma_{MD'[(i) } \sigma_{NO'(j)]}\epsilon^{D'K'}\!\epsilon^{O'S'} \!\epsilon^{MN} J_{S'K'A'}   }
%\end{eqnarray}

\subsubsection{The effects of the perturbed tetrad basis}
Now we analyze $(I_1)_{(\tau)(\beta)(\gamma)A'}$. Since it has the full anti-symmetrization for the indices $(\tau),(\beta),(\gamma)$, i.e., $(I_1)_{(i)(j)(k)A'}$ belongs to the fundamental spinor representation of SO(3) group, thus it cannot make contribution to the scattering of the massless wave with spin $\frac{3}{2}$. So we focus on $(I_1)_{(i)(j)(0)A'}$, which can be written as
\begin{eqnarray}
&&(I_1)_{(i)(j)(0)A'}=-\frac{\text{i}M}{6r^3}x^n  k_{(\text{in})n } \sigma_{BA'(0) } \xi_{(\text{in})}^B (x_i  m_{(j) }-x_j  m_{(i) } )\text{e}^{\text{i} k_{(\text{in})\alpha} x^\alpha}\nonumber\\
&& ~~~~ +\frac{\text{i}M}{3r}\omega \sigma_{BA'[(i) } m_{(j)] }  \xi_{(\text{in})}^B \text{e}^{\text{i} k_{(\text{in})\alpha} x^\alpha}~.\label{T8}
\end{eqnarray}
Since
\begin{eqnarray}
&&x_i  m_{(j) }-x_j  m_{(i) }=-\frac{x_m}{2\sqrt2} \sigma_{MT'[(i) } \sigma_{NO'(j)]} \sigma_{IJ'}^{(0) } \epsilon^{MN }\epsilon^{J'H'}  \epsilon^{Q'T'} \epsilon^{S'O'}\nonumber\\
&&~~~~ \times   \Big(\epsilon^{CE} \sigma_{EH'}^{(m) } \sigma_{CQ'}^{(0)} \sigma_{GS'}^{(0)}
-\epsilon_{H'S' }\sigma_{GQ'}^{(m) } \Big) \xi_{(\text{in})}^G \xi_{(\text{in})}^I~,
\end{eqnarray}
we have
\begin{eqnarray}
&&\hskip -0.75cm (I_1)_{(i)(j)(0)A'}
=\frac{\text{i}M\omega}{3r} \sigma_{BA'[(i) } m_{(j)] } \xi_{(\text{in})}^B \text{e}^{\text{i} k_{(\text{in})\alpha} x^\alpha} \nonumber\\
&& \hskip -0.75cm ~~~~ + \,\frac{\text{i}M x^n k_{(\text{in})n }}{6\sqrt2 r^3}\sigma_{MD'[(i) } \sigma_{NO'(j)]} \epsilon^{D'K'} \!\epsilon^{O'S'}   \!\epsilon^{MN} x_l  \sigma_{EK'}^{ (l) }  \sigma_{GS'}^{ (0) } \sigma_{BA'}^{ (0) } \xi_{(\text{in})}^E\xi_{(\text{in})}^G \xi_{(\text{in})}^B\text{e}^{\text{i} k_{(\text{in})\alpha} x^\alpha}\nonumber\\
&& \hskip -0.75cm ~~~~ +\, \frac{\text{i}M x^n k_{(\text{in})n }}{6\sqrt2 r^3}\sigma_{MD'[(i) } \sigma_{NO'(j)]} \epsilon^{D'K'} \!\epsilon^{O'S'}   \! \epsilon^{MN}  [\epsilon_{K'A'} (B_5)_{S'}\!+\!\epsilon_{S'A'} (B_6)_{K'} ]\text{e}^{\text{i} k_{(\text{in})\alpha} x^\alpha},\label{hskip -}
%\nonumber\\
%&& \hskip -0.75cm~,
\end{eqnarray}
where $(B_5)_{S'}$ and $(B_6)_{K'}$ are complex spinors with spin $\frac{1}{2}$, thus they do not contribute to the scattering of the massless wave with spin $\frac{3}{2}$. Taking Fourier expansion for $\frac{x^n}{r^3}$ in the second term, we have
\begin{eqnarray}
&&\hskip -1cm (I_1)_{(i)(j)(0)A'}=\frac{\text{i}M\omega}{3r}  \sigma_{BA'[(i) } m_{(j)] }  \xi_{(\text{in})}^B \text{e}^{\text{i} k_{(\text{in})\alpha} x^\alpha}+\frac{\text{i}Mk_{(\text{in})}^n}{12\pi^2 \sqrt2}  \sigma_{MD'[(i) } \sigma_{NO'(j)]} \nonumber\\
 &&\hskip -1cm~~~~\times  \epsilon^{D'K'} \!\epsilon^{O'S'} \!\epsilon^{MN} \sigma_{EK'}^{(l) }  \sigma_{GS'}^{(0)} \sigma_{BA'}^{ (0) }  \xi_{(\text{in})}^E \xi_{(\text{in})}^G \xi_{(\text{in})}^B \!\int\! \Big(\frac{\delta _{nl}}{\bf{k}^2_{\text{(tran)}}}\!-\!2\frac{k_{\text{(tran)}n} k_{\text{(tran)}l}}{\bf{k}^4_{\text{(tran)}}}\Big)\text{e}^{\text{i} k_{(\text{out})\alpha} x^\alpha} d^3k_{\text{(tran)}} \nonumber\\
&&\hskip -1cm~~~~+\,\frac{\text{i}Mx^n k_{(\text{in})n}}{6\sqrt2 r^3} \sigma_{MD'[(i) } \sigma_{NO'(j)]}   \epsilon^{D'K'}\! \epsilon^{O'S'} \! \epsilon^{MN} [\epsilon_{K'A'} (B_5)_{S'}+\epsilon_{S'A'} (B_6)_{K'}]\text{e}^{\text{i} k_{(\text{in})\alpha} x^\alpha},\label{hskip -0}
\ffa
where $k_{(\text{out})\alpha}\!=\!k_{(\text{tran})\alpha}\!+\!k_{(\text{in})\alpha}$ is the outgoing wave vector. Here, the first term contains $\sigma_{BA'[(i) } m_{(j)] } \xi_{(\text{in})}^B$, which after tedious calculation can be expressed as
\begin{eqnarray}
&&\sigma_{BA'[(i) }^{~} m_{(j)] }^{~} \xi_{(\text{in})}^B = \frac{\sigma_{MD'[(i) } \sigma_{NO'(j)] }}{2\sqrt2} \epsilon^{MN}\! \epsilon^{D'K'} \! \epsilon^{O'S'}[\sigma_{BK'}^{ (0) } \sigma_{GS'}^{  (0) } \sigma_{IA'}^{(0) }\xi_{(\text{in})}^G \xi_{(\text{in})}^I \xi_{(\text{in})}^B\nonumber \!+\!\epsilon_{K'A'} (B_7)_{S'}],
\end{eqnarray}
with $(B_7)_{S'}$  being a complex spinor with spin $\frac{1}{2}$ which cannot contribute to the scattering of the massless wave with spin $\frac{3}{2}$. Therefore, we can write $(I_1)_{(i)(j)(0)A'}$ as
\begin{eqnarray}
&&\hskip -0.75cm (I_1)_{(i)(j)(0)A'}=\frac{\text{i}M}{12\pi^2 \sqrt2}  \sigma_{MD'[(i) } \sigma_{NO'(j)]}    \epsilon^{D'K'} \epsilon^{O'S'} \epsilon^{MN} \sigma_{GS'}^{(0)}   \sigma_{BA'}^{ (0) } \xi_{(\text{in})}^E\xi_{(\text{in})}^G \xi_{(\text{in})}^B \nonumber\\
&&\hskip -0.75cm ~~~~\times \Big[ k_{(\text{in})}^n  \sigma_{EK'}^{(l) } \!\!\int\!   \Big(\delta _{nl}\!-\!2\frac{k_{\text{(tran)}n} k_{\text{(tran)}l}}{\bf{k}^2_{\text{(tran)}}}\Big) \text{e}^{\text{i} k_{(\text{out})\alpha} x^\alpha} \frac{d^3k_{\text{(tran)}}}{\bf{k}^2_{\text{(tran)}}}  + \omega  \sigma_{EK'}^{(0)} \!\!\int\! \text{e}^{\text{i} k_{(\text{out})\alpha} x^\alpha}  \frac{d^3k_{\text{(tran)}}}{\bf{k}^2_{\text{(tran)}}} \Big]\nonumber\\
&&\hskip -0.75cm ~~~~+\,\sigma_{MD'[(i) } \sigma_{NO'(j)]}    \epsilon^{D'K'} \epsilon^{O'S'} \epsilon^{MN} \Big[\epsilon_{K'A'} (B_8)_{S'}+\frac{\text{i}Mx^n  k_{(\text{in})n}}{6\sqrt2 r^3} \epsilon_{S'A'} (B_6)_{K'}\Big]~,\label{hski}
\end{eqnarray}
where $(B_8)_{S'}\!\equiv\!\frac{\text{i}M}{6\sqrt2 }\Big[ \frac{x^nk_{(\text{in})n}}{r^3}(B_5)_{S'}+\frac{\omega}{r}(B_7)_{S'}\big]\text{e}^{\text{i} k_{(\text{in})\alpha} x^\alpha}$ is a complex spinor with spin $\frac{1}{2}$, thus it do not contribute to the scattering of the massless wave with spin $\frac{3}{2}$. We can re-write $(I_1)_{(i)(j)(0)A'}$ as follow
\begin{eqnarray}
&&\hskip -0.75cm (I_1)_{(i)(j)(0)A'}= 2\sigma_{MD'[(i) } \sigma_{NO'(j)]}  \epsilon^{D'K'}\! \epsilon^{O'S'}\! \epsilon^{MN} \!\!\int\!\! (J_1 )_{S'K'A'} \exp\{\text{i}(k_{\text{(tran)}\alpha}\!+\!k_{(\text{in})\alpha} ) x^\alpha \}d^3k_{\text{(tran)}} \nonumber\\
&&\hskip -0.75cm ~~~~ +\,2\sigma_{MD'[(i) } \sigma_{NO'(j)]}    \epsilon^{D'K'} \epsilon^{O'S'} \epsilon^{MN} \Big[\epsilon_{K'A'} (B_9)_{S'}+\epsilon_{S'A'} (B_{10})_{K'}\Big]~,\label{-}
\end{eqnarray}
with
\begin{eqnarray}
&&\hskip -0.75cm (J_1 )_{S'K'A'}=\frac{\text{i}M}{72\sqrt2 \pi^2}\xi_{(\text{in})}^E\xi_{(\text{in})}^G\xi_{(\text{in})}^B\Big[
\frac{3\omega}{\bm{k}^2_{\text{(tran)}}}\sigma_{BK'}^{ (0) } \sigma_{GS'}^{(0)} \sigma_{EA'}^{(0) } \nonumber\\
&& \hskip -2.75cm ~~~~\,+
\Big(\sigma_{GS'}^{(0)} \sigma_{EK'}^{(l)}  \sigma_{BA'}^{ (0) }  +\sigma_{GK'}^{(0)} \sigma_{EA'}^{(l)}  \sigma_{BS'}^{ (0) }  +\sigma_{GA'}^{(0)} \sigma_{ES'}^{(l)}  \sigma_{BK'}^{ (0) }   \Big)\Big(\frac{k_{(\text{in})l}}{\bm{k}^2_{\text{(tran)}}}-2\frac{k_{(\text{in})}^n  k_{\text{(tran)}n}  k_{\text{(tran)}l}}{\bm{k}^4} \Big)\Big]~,\label{-0}
\end{eqnarray}
where $(B_9)_{S'}$ and $(B_{10})_{K'}$ are complex spinors with spin $\frac{1}{2}$, thus they do not contribute to the scattering of the massless wave with spin $\frac{3}{2}$.

Finally, we can obtain the contribution of $(I_1)_{(i)(j)(0)A'}$ to the massless wave with spin $\frac{3}{2}$ as follow
\fa
2 \sigma_{MD'[(i) } \sigma_{NO'(j)]}\epsilon^{MN}\!\epsilon^{D'K'}\!\epsilon^{O'S'} \! (J_1 )_{S'K'A'} ~.\label{s}
\ffa

\section{Solution based on the simulating wave function method}

Inspired by Archimedes’ method for calculating buoyancy, we simulate a wave function to compute the scattering amplitude of gravitino in the Schwarzschild field. We then estimate the discrepancy between the simulated wave function and the true one, and demonstrate that this difference is negligible.

\subsection{The method of simulating wave function}
We assume scattering occurs instantaneously and scattered wave propagates in some direction. So we need construct some bases.

First, setting
\fa
\psi_{\text{scat}(\tau)}^B(x^{j},t,t_0)\equiv H(t-t_0)\iota_{(\tau)}^B \text{e}^{\text{i} k_{(\text{sim})\alpha} x^\alpha}~,\label{psis}
\ffa
where $t_0$ denotes the time when the scattering happens. $H(s)$ is the step function of $s$. $k_{(\text{sim})\alpha}$ denotes the wave vector of the simulating wave function. $\iota_{(\tau)}^B$ denotes the polarization of the scattering wave and it is the function of $\bm{k}_{(\text{sim})}$.
Since the helicity of the outgoing wave is $\frac{3}{2}$, $\iota _{(\tau)}^B$ satisfies
\fa
\sigma _{BA'[(\gamma) }^{~} k_{(\text{sim})(\beta)}^{~} \iota _{(\tau)]}^B=0~,\label{iota}
\ffa
with
\fa
&& k_{(\text{sim})\alpha}^{~}k_{(\text{sim})}^\alpha=0~,\label{kk}\\
&& k_{(\text{sim})0} \le 0~.
\ffa

Similar to the equations-of-motion for the gravitino wave, see Eq. (\ref{psiEoB}), we construct $(I_{<1>})_{(\tau)(\beta)(\gamma)A'}$ to satisfy
\fa
%\hskip -0.5cm
\sigma _{BA'[(\tau) } ^{~}\partial _\beta \psi_{\text{scat}(\gamma)]}^B +\sigma _{BA'[(\beta) }^{~} \partial _\gamma \psi_{\text{scat}(\tau)]}^B+\sigma _{BA'[(\gamma) } ^{~} \partial _\tau \psi_{\text{scat}(\beta)]}^B +3(I_{<1>})_{(\tau)(\beta)(\gamma)A'} =0.\label{psisa}
\ffa

Substituting Eq.\,(\ref{psis}) into Eq.\,(\ref{psisa}), making use of Eq.\,(\ref{iota}), we have

\begin{eqnarray}
&& (I_{<1>})_{(i)(j)(0)A'} (t)%x^{b}\}\nn\\&& ~~~~
=-\frac{1}{3}\sigma_{BA'[(j)}^{~} \iota_{(i)]}^B \delta(t-t_0)\text{e}^{\text{i} k_{(\text{sim})l} x^l}
\text{e}^{-\text{i} \omega_{sim} t_0},\label{I_}
\end{eqnarray}
with $\omega_{sim}\!=\!-k_{(\text{sim})0}$.\\

Let
\begin{eqnarray}
\Psi_{(\tau)}^B\equiv\mu_{(\tau)}^B\text{e}^{\text{i} k_{(\text{sim})l} x^l}\text{e}^{-\text{i} \omega t}~,\label{Psi}
\end{eqnarray}
with $\mu_{(\tau)}^B$ being a spin-vector, to simulate the changes in the bounded state of gravitino caused by gravitational field. Then we construct $I_{<2>}$ to satisfy
\fa
&&\sigma _{BA'[(\tau) }^{~} \partial _\beta \Psi_{(\gamma)]}^B+\sigma _{BA'[(\beta) }^{~} \partial _\gamma \Psi_{(\tau)]}^B+\sigma _{BA'[(\gamma) }^{~} \partial _\tau \Psi_{(\beta)]}^B+3(I_{<2>})_{(\tau)(\beta)(\gamma)A'}=0~.\label{I49}
\ffa

Plugging Eq.\,(\ref{Psi}) into Eq.\,(\ref{I49}), we have
\begin{eqnarray}
&& \hskip -1.5cm (I_{<2>})_{(i)(j)(0)A'}= -\frac{\text{i}}{6}\big[\sigma _{BA'(0) } \big(k_{\text{(sim)}i}  \mu _{(j)}^B-k_{\text{(sim)}j}  \mu _{(i)}^B\big) +2\omega \sigma _{BA'[(i) }  \mu _{(j)]}^B\big]\text{e}^{\text{i} k_{(\text{out})l} x^l}\text{e}^{-\text{i} \omega t},    \label{I<2>}
\end{eqnarray}

with
\fa
&& \hskip -0.5cm \sigma _{BA'[(i) }  \mu _{(j)]}^B
=\frac{1}{2}\sigma _{MD'[(i) } \sigma _{NO'(j)]}  \epsilon ^{MN} \epsilon ^{O'S'} \big(\sigma _{BK'}^{(0)} \epsilon ^{K'D'} \sigma _{GS'}^{(0)} \epsilon ^{CG} \sigma _{CA'}^{(k) }+\delta^{D'}_{A'}\sigma _{BS'}^{(k) } \big) \mu _{(k)}^B\nn\\
&&~~~~~~~~~~~\, =-\frac{1}{4} \sigma _{MD'[(i) } \sigma _{NO'(j)]} \epsilon ^{MN} \epsilon ^{O'S'} \epsilon ^{D'K'}
\nn\\
&&~~~~~~~~~~~~~~~~\times \big[\epsilon ^{CG} \sigma _{BK'}^{(0)} \sigma _{GS'}^{(0)} \sigma _{CA'}^{(k) } \mu _{(k)}^B
+2\big(\epsilon _{K’A'} (B_{11} )_{S'}+\epsilon _{S’A'} (B_{12} )_{K'} \big)\big]~,\label{az}
\ffa
\fa
&&  \hskip -1cm  k_{\text{(sim)}i}  \mu _{(j)}^B-k_{\text{(sim)}j}  \mu _{(i)}^B = -\frac{1}{2} k_{\text{(sim)}l}\sigma_{IK'}^l\epsilon^{IG}\epsilon^{K'S'} 
(\sigma _{GS'(i) } \mu _{(j)}^B - \sigma _{GS'(j) } \mu _{(i)}^B) \nn\\
&&~~~~~~~~\,~~~=\frac{1}{2}\sigma _{MD'[(i) } \sigma _{NO'(j)]}  \epsilon ^{MN} \epsilon ^{O'S'} \epsilon^{D'K'}k_{\text{(sim)}l}\sigma _{IK'}^{(l) }\epsilon^{IG} \sigma _{GS'}^{(k) }    \mu _{(k)}^B~,\label{aa}
\ffa
%where $(B_{13} )_{S'}$ and $(B_{14} )_{S'}$ do not contribute to the scattering of gravitino wave.
where $(B_{11})_{S'}$ and $(B_{12})_{K'}$ are complex spinors with spin $\frac{1}{2}$, thus they do not contribute to the scattering of the massless wave with spin $\frac{3}{2}$.

For calculating $\frac{d\sigma}{d\Omega}|_{M\omega\rightarrow 0} $, we need to find the relation between the cross section and the first-order perturbation term of the total wave. This can be achieved in two steps:
(1) We establish the relationship between the simulation coefficients and the true wave function, which can be derived from analyzing the degree of freedom of every helicities. The simulation coefficients will satisfy  a variational extremal condition.
(2) We establish the relationship between the scattering cross-section and the simulation coefficients. First, we construct a scattering wave in an arbitrary direction. Second, we perform a linear superposition of scattering waves from all directions. Third, by calculating the number of gravitational neutrinos scattered per unit time in each direction, we determine the relationship between the weight of each directional scattering wave and the corresponding scattering cross-section.

These two steps will be dealt with in the following two subsections, respectively.

\subsection{Relation between the simulation coefficients and the true wave function}

We integrate these bases to simulate the real outgoing wave denoted by $\psi _{\text{sim}(\tau)}^B$.

Since the scattering happens at all time and in all direction, we need to do the following integration
\fa
\psi _{\text{sim}(\tau)}^B(x^j,t)\equiv\int _{-T/2}^{T/2}dt_0\int \psi_{\text{scat}(\tau)}^B(x^j,t,t_0)\text{e}^{\text{i}(\omega_{sim} -\omega )t_0}d^3 k_{(\text{sim})}  ~.\label{ab}
\ffa
Notice that $\psi_{\text{scat}(\tau)}^B$ is given in Eq.\,(\ref{psis}).

Similar to the equations-of-motion for the gravitino wave, see Eq. (\ref{psiEoB}), we construct $(I_{<3>})$ to satisfy
\fa
&& \sigma _{BA'[(\tau) }^{~} \partial _\beta \psi _{\text{sim}(\gamma)]}^B+\sigma _{BA'[(\beta) }^{~} \partial _\gamma \psi _{\text{sim}(\tau)]}^B+\sigma _{BA'[(\gamma) }^{~} \partial _\tau \psi _{\text{sim}(\beta)]}^B\nn+3(I_{<3>})_{(\tau)(\beta)(\gamma)A'}=0~,\\\label{sigmapsi}
\ffa
then we have
\fa
&&(I_{<3>})_{(i)(j)(k)A'}=0~,\label{ad}\\
&&(I_{<3>})_{(i)(j)(0)A'}=\frac{1}{3}\int\!  \sigma_{BA'[(i) }^{~} \iota _{(j)]}^B \text{e}^{\text{i} k_{(\text{sim})l} x^l}\text{e}^{-\text{i} \omega t}d^3 k_{(\text{sim})}~,\label{ae}
\ffa

with
\fa
&& \sigma _{BA'[(i) } \iota _{(j)]}^B=\frac{\sigma _{MD'[(i) } \sigma _{NO'(j)]} }{4} \epsilon ^{MN} \epsilon ^{O'S'} (\sigma _{BK'}^{(0)} \epsilon ^{K'D'} \sigma _{GS'}^{(0)} \epsilon ^{CG} \sigma _{CA'}^{(k) }+\delta^{D'}_{A'}\sigma _{BS'}^{(k) } ) \iota _{(k)}^B\nn\\
&&~~~~~~~~=-\frac{1}{4} \sigma _{MD'[(i) } \sigma _{NO'(j)]} \epsilon ^{MN} \epsilon ^{O'S'} \epsilon ^{D'K'} \epsilon ^{CG} \sigma _{BK'}^{(0)} \sigma _{GS'}^{(0)} \sigma _{CA'}^{(k) } \iota _{(k)}^B\nn\\
&&~~~~~~~~~~~~~-\sigma _{MD'[(i) } \sigma _{NO'(j)]} \epsilon ^{MN} \epsilon ^{O'S'} \epsilon ^{D'K'} \big[\epsilon _{K’A'} (B_{13} )_{S'}+\epsilon _{S’A'} (B_{14} )_{K'} \big]~.\label{sigio}
\ffa
Therefore,
\fa
&& (I_{<3>})_{(i)(j)(0)A'}
=\frac{1}{12}\!\int\! d^3 k_{(\text{sim})} \sigma _{MD'[(i) } \sigma _{NO'(j)]} \epsilon ^{MN} \epsilon ^{O'S'} \epsilon ^{D'K'}\text{e}^{\text{i} k_{(\text{sim})l} x^l-\text{i} \omega t}\nn\\
&& ~~~~ \times [ \epsilon ^{CG} \sigma _{BK'}^{(0)} \sigma _{GS'}^{(0)} \sigma _{CA'}^{(k) } \iota _{(k)}^B  +4 \epsilon _{K'A'} (B_{13} )_{S'}+4\epsilon _{S’A'} (B_{14} )_{K'} ]~,\label{ag}
\ffa
where $(B_{13})_{S'}$ and $(B_{14})_{K'}$ are complex spinors with spin $\frac{1}{2}$, thus they do not contribute to the scattering of the massless wave with spin $\frac{3}{2}$.

Let $\Psi_{sim{(\tau)}}^B$ simulate the bounded part of the gravitino wave from all direction, we have
\begin{eqnarray}
\Psi_{sim{(\tau)}}^B=\int \Psi_{(\tau)}^B d^3 k_{(\text{sim})}  ~. \label{sim}
\end{eqnarray}
Similar to the equations-of-motion for the gravitino wave, see Eq. (\ref{psiEoB}), we construct $(I_{<4>})$ to satisfy
\fa
\sigma _{BA'[(\tau) }^{~} \partial _\beta \Psi _{\text{sim}(\gamma)]}^B+\sigma _{BA'[(\beta) }^{~} \partial _\gamma \Psi _{\text{sim}(\tau)]}^B+\sigma _{BA'[(\gamma) }^{~} \partial _\tau \Psi _{\text{sim}(\beta)]}^B\nn +3(I_{<4>})_{(\tau)(\beta)(\gamma)A'}=0,\label{sigmaPsi}
\ffa
then we have
\begin{eqnarray}
(I_{<4>})_{(i)(j)(0)A'}=\int (I_{<2>})_{(i)(j)(0)A'} d^3 k_{(\text{sim})}~, \label{I<4>}
\end{eqnarray}

Substituting Eq.\,(\ref{I<2>}) into Eq.\,(\ref{I<4>}), we have
\begin{eqnarray}
&& (I_{<4>})_{(i)(j)(0)A'}
=\frac{\text{i}}{6}\sigma _{MD'[(i) } \sigma _{NO'(j)]}  \epsilon ^{MN} \epsilon ^{O'S'} \epsilon^{D'K'}\nonumber\\
&& ~~~~ \times \int d^3 k_{(\text{sim})}\Big[-\frac{1}{2}\sigma _{BA'(0) }k_{\text{(sim)}l}\sigma _{EK'}^{(l) }\epsilon^{EG} \sigma _{GS'}^{(k) }    \mu _{(k)}^B+\frac{\omega}{2}\epsilon ^{CG} \sigma _{BK'}^{(0)} \sigma _{GS'}^{(0)} \sigma _{CA'}^{(k) } \mu _{(k)}^B\nonumber\\
&& ~~~~~~~~~~~~
+\,\omega\big(\epsilon _{K’A'} (B_{11} )_{S'}+\epsilon _{S’A'} (B_{12} )_{K'} \big)\Big]
\text{e}^{\text{i} k_{(\text{sim})l}x^l}\text{e}^{-\text{i} \omega t}~.\label{bz}
\end{eqnarray}
%where $(B_{11})_{S'}$ and $(B_{12})_{K'}$ are complex spinors with spin $\frac{1}{2}$, thus they do not contribute to the scattering of the massless wave with spin $\frac{3}{2}$.

Finally, we can obtain
\begin{eqnarray}
&& \hskip -1.cm (I_{<3>})_{(i)(j)(0)A'}+(I_{<4>})_{(i)(j)(0)A'}
=\sigma _{MD'[(i) } \sigma _{NO'(j)]} \epsilon ^{MN} \epsilon ^{O'S'} \epsilon ^{D'K'}
\nonumber\\
&&  \hskip -1cm ~~~~~~\times \Big\{\frac{1}{12} \int\! d^3k_{(\text{sim})}
\epsilon ^{CG} \sigma _{BK'}^{(0)} \sigma _{GS'}^{(0)}
\sigma _{CA'}^{(k)} \big(\iota _{(k)}^B+\text{i}\omega\mu_{(k)}^B\big)\text{e}^{\text{i} k_{(\text{sim})l} x^l}\text{e}^{-\text{i} \omega t}
\nn\\
&&  \hskip -1cm ~~~~~~~~~~
-\frac{\text{i}}{12} \int\! d^3k_{(\text{sim})} \sigma_{BA'(0) }k_{(\text{sim})l}\sigma _{EK'}^{(l) }\epsilon^{EG} \sigma _{GS'}^{(k)} \mu _{(k)}^B\text{e}^{\text{i} k_{(\text{sim})l} x^l}\text{e}^{-\text{i} \omega t}\nonumber\\
&&  \hskip -1cm ~~~~~~~~~~
+\int\! d^3 k_{(\text{sim})}\big(\epsilon _{K’A'} (B_{15} )_{S'}+\epsilon _{S’A'} (B_{16} )_{K'}\big)\text{e}^{\text{i} k_{(\text{sim})l} x^l}\text{e}^{-\text{i} \omega t}\Big\}~,\label{ba}
\end{eqnarray}
where $(B_{15})_{S'}$ and $(B_{16})_{K'}$ are complex spinors with spin $\frac{1}{2}$, thus they do not contribute to the scattering of the massless wave with spin $\frac{3}{2}$.

In order to calculate the differential cross-section at a generic direction, we draw a small neighbourhood $\Theta$ around the outgoing wave vector $k_{(\text{out})l}$ without $0$ for excluding the issue of a zero denominator. Then according to the consistency of equations-of-motion for the gravitino wave, we can adjust $\iota _{(k)}^B $ and $\mu _{(k)}^B $ in $k\in \Theta$ to meet
\begin{eqnarray}
&&  (J_1)_{S'K'A'}+J_{S'K'A'} = \frac{1}{24}\epsilon ^{CG} (\sigma _{B(K'}^{(0)} \sigma _{GS'}^{(0)} \sigma _{CA')}^{(j) }
\big( \iota _{(j)}^B+\text{i} \omega \mu _{(j)}^B\big)\nn\\
&& ~~~~
+\frac{\text{i}k_{(\text{out})l}}{24}\epsilon^{GC}  \sigma _{B(K'}^{(0)} \sigma _{GS'}^{(l) }\sigma _{CA')}^{(j)} \mu_{(j)}^B~. \label{goal}
\end{eqnarray}
This concludes the second step.

\subsection{Calculation of cross-section}

We first calculate the cross-section of the simulation wave, and then demonstrate that the scattering amplitude of the simulated wave and that of the real wave are the same. This section is divided into three sub-sections.

\subsubsection{Decomposition of true wave function}
Since the simulated waves $\psi_{\text{sim}(\gamma)}^B$ and $\Psi_{\text{sim}(\gamma)}^B$ both contain the component of the spin-$\frac{1}{2}$ wave, we need to exclude it from the scattering of the gravitino wave
\fa
\psi_{\{1\}(\gamma)}^B =  \psi_{\text{sim}(\gamma)}^B+\Psi_{\text{sim}(\gamma)}^B-\psi_{\{1\}(\gamma)}^{(s=\frac{1}{2})B} ~.\label{psi1cB}
\ffa

\subsubsection{Relation between cross-section and simulating coefficient}
We calculate the scattering amplitude in $\psi_{\text{sim}(\gamma)}^B$. In order to do this, we need the particle current of scattering wave in all directions, which can be obtained from the algebra relation between the particle number and the gravitino field.

For the plane wave constructed in Eq.\,(\ref{z4}), we assume the particle density is $\varepsilon ^2 \rho $. According to fundamental principle of quantum field theory, the number of particles is a quadratic form of the field intensity.
For a monochromatic wave, theoretically, any quadratic form is acceptable. However, in practical situations, gravitational wavelets have a spatial extent. Therefore, the particle number should be expressed as an integral over the entire space. It is also acceptable to write it simply as
\fa
N_{\text{one~direction}}=\frac{\rho'}{2} \int d^3 x |
\varepsilon \bar{m}_{(\tau)} \xi ^B |^2~,\label{bd}
\ffa
But it is equivalent to
\fa
N_{\text{one~direction}}=\frac{\rho}{2} \int d^3 x |\epsilon ^{CG} \sigma _{CD'}^{(0)} \sigma _{BF'}^{(0)} (\xi ^* )^{D'} (\xi ^* )^{F'} (\xi ^* )^{S'} \sigma _{GS'}^{(\tau)}
\varepsilon \bar{m}_{(\tau)} \xi ^B |^2~,
\ffa
here the coefficient is $\rho/2$ is determined based on the previously defined density as follows:
First we make use of the identity
\fa
-\sqrt{2} \varepsilon \equiv - \sigma _{\uparrow \downarrow '}^{(\tau) } \varepsilon \bar{m}_{(\tau)}
 \equiv
\epsilon ^{CG} \sigma _{CD'}^{(0)} \sigma _{AF'}^{(0)}
(\xi ^* )^{D'} (\xi ^* )^{F'} (\xi ^* )^{S'}
\sigma _{GS'}^{(\tau) } \varepsilon \bar{m}_{(\tau)} \xi ^A ~.
\ffa

Therefore, the number of particles of the system for single wavelength along one direction can be written as
\fa
N_{\text{one~direction}}=\frac{\rho}{2} \int d^3 x |\epsilon ^{CG} \sigma _{CD'}^{(0)} \sigma _{BF'}^{(0)} (\xi ^* )^{D'} (\xi ^* )^{F'} (\xi ^* )^{S'} \sigma _{GS'}^{(\tau)}
\varepsilon \bar{m}_{(\tau)} \xi ^B |^2~.
\ffa

Let
\fa
\psi_{\text{all~direction}(\tau)}^B \equiv \int   \psi _{\text{scat}(\tau)}^B|_{t_0=0} ~d^3 k_{(\text{sim})}~.
\ffa
We assume that there is a $\delta \omega$ such that $0<\delta \omega<<\omega$ and either $|\omega_{sim}-\omega|<<\delta \omega$ or $\iota_{(\tau)}^B\approx 0$,
then the number of particles from all directions for $\psi_{\text{all~direction}(\tau)}^B$ can be written as
\fa
N_{\text{all~direction}}=4\pi ^3 \rho  \int d^3 k_{(\text{sim})}  |\epsilon ^{CG} \sigma _{CD'}^{(0)} \sigma _{BF'}^{(0)} (\xi _{sim}^* )^{D'} (\xi _{sim}^* )^{F'} (\xi _{sim}^* )^{S'} \sigma _{GS'}^{(\tau) } \iota _{(\tau)}^B |^2~.
\ffa
For $\psi _{\text{sim}(\tau)}^B$,  when $t>T/2$,  number of particles
\begin{eqnarray}
&&N_{\text{all~direction}} =4\pi ^3 \rho \int d^3 k_{(\text{sim})}  \Big[|\epsilon ^{CG} \sigma _{CD'}^{(0)} \sigma _{BF'}^{(0)} (\xi _{sim}^* )^{D'} (\xi _{sim}^* )^{F'} (\xi _{sim}^* )^{S'} \sigma _{GS'}^{(\tau) } \iota _{(\tau)}^B|^2\nonumber\\
&&~~~~\hskip 1cm \times \Big|\int _{-T/2}^{T/2}dt_0 \text{e}^{\text{i}(\omega_{sim}-\omega )t_0}\Big|^2 \Big]\nonumber\\
&&~~~~ =4\pi ^3 \rho \int d^3 k_{(\text{sim})}   \Big[\big|\epsilon ^{CG} \sigma _{CD'}^{(0)} \sigma _{BF'}^{(0)} (\xi _{sim}^* )^{D'} (\xi _{sim}^* )^{F'} (\xi _{sim}^* )^{S'} \sigma _{GS'}^{(\tau) } \iota _{(\tau)}^B \big|^2\nonumber\\
&&~~~~\hskip 1cm \times \Big|\frac{2\sin [(\omega_{sim}-\omega )\frac{T}{2}] } {\omega_{sim}-\omega}\Big|^2  \Big]~\nn\\
&&~~~~=   8\pi ^4 T\rho  \int d^3 k_{(\text{sim})}  \Big[\big|\epsilon ^{CG} \sigma _{CD'}^{(0)} \sigma _{BF'}^{(0)} (\xi _{sim}^* )^{D'} \!(\xi _{sim}^* )^{F'} \!(\xi _{sim}^* )^{S'} \sigma _{GS'}^{(\tau) } \iota _{(\tau)}^B\big|^2 \delta (\omega_{sim}\!-\!\omega )\Big]\nn\\
&&~~~~ =  8\pi ^4 T\rho  \omega^2 \int \sin\theta' d\theta' d\phi'  \big|\epsilon ^{CG} \sigma _{CD'}^{(0)} \sigma _{BF'}^{(0)} (\xi _{sim}^* )^{D'} (\xi _{sim}^* )^{F'} (\xi _{sim}^* )^{S'} \sigma _{GS'}^{(\tau) } \iota _{(\tau)}^B \big|^2~,\label{cz}
\end{eqnarray}
where we have used $\big|\frac{2\sin(\frac{y}{2})}{y}\big|^2 \approx 2\pi \delta (y)$ for $y >> 1$. 
%And $\bf{n}(\theta,\phi)=\sin\theta\cos\phi\bf{e}_x+\sin\theta\sin\phi\bf{e}_y+\cos\theta\bf{e}_z$.\\
%The relation between $(\xi _{sim}^* )^{D'}$ and $(\theta',\phi')$ is\\
$(\xi _{sim}^* )^{\uparrow'}=-\sin(\theta'/2)\text{e}^{\text{i}\phi'/2}$, and $(\xi _{sim}^* )^{\downarrow'}=-\cos(\theta'/2)\text{e}^{-\text{i}\phi'/2}$. For the solid angle $(\theta, \phi)$, the differential current of particles induced by $\psi_{\text{sim}(\gamma)}^B$ can be written as
\fa
&& \lim_{M\omega\rightarrow 0}  \frac{dP^{(\psi_{\text{sim}(\gamma)}^B)}}{d\Omega} =8\pi ^4 \rho \omega ^2
\big|\epsilon ^{CG} \sigma _{CD'}^{(0)} \sigma _{BF'}^{(0)} (\xi _{sim}^* )^{D'} (\xi _{sim}^* )^{F'} (\xi _{sim}^* )^{S'} \sigma _{GS'}^{ (\tau) } \iota _{(\tau)}^B \big|^2~\nn\\
&& ~~~~ =8\pi ^4 \rho \omega^2 \big|\epsilon ^{CG} \sigma _{CA'}^{(0)} \sigma _{BK'}^{(0)} (\xi _{out}^* )^{A'} (\xi _{out}^* )^{K'} (\xi _{out}^* )^{S'} \sigma _{GS'}^{(\tau) } \iota _{(\tau)}^B \big|^2~.
\ffa
where we have taken into account of the spherical symmetry. %Therefore, 
  
%Without loss of generality, let $\phi =0$,then $(\xi _{sim}^* )^{S'}(\theta,\phi)=(\xi _{out}^* )^{S'}$.\\
Since the current density of particles of the incident wave is $\rho $,  the differential cross-section contributed by $\psi_{\text{sim}(\gamma)}^B$ is
\fa
\lim_{M\omega\rightarrow 0} \frac{d\sigma^{(\psi_{\text{sim}(\gamma)}^B)}}{d\Omega}   =8\pi ^4 \omega^2 \big|\epsilon ^{CG} \sigma _{CA'}^{(0)} \sigma _{BK'}^{(0)} (\xi _{out}^* )^{A'} (\xi _{out}^* )^{K'} (\xi _{out}^* )^{S'} \sigma _{GS'}^{(\tau) } \iota _{(\tau)}^B \big|^2~.\label{cb}
\ffa

From \eqref{goal}, at the sphere $|k_{sim}|=\omega$, we have
\fa
&&\epsilon ^{GC} (\xi _{out}^* )^{S'} (\xi _{out}^* )^{K'} (\xi _{out}^* )^{A'} \sigma _{BK'}^{(0)} \sigma _{CA'}^{(0)} \sigma _{GS'}^{(k) } \iota _{(k)}^B\nn \\
&&~~~~=24 (\xi _{out}^* )^{A'} (\xi _{out}^* )^{K'} (\xi _{out}^* )^{S'} (J_{S'K'A'}+(J_1)_{S'K'A'})~,\label{cc}
\ffa
therefore, 
\begin{eqnarray}
 \lim_{M\omega\rightarrow 0} \frac{d\sigma^{(\psi_{\text{sim}(\gamma)}^B)}}{d\Omega} = 4608\pi ^4\omega^2 |(\xi _{out}^* )^{A'} (\xi _{out}^* )^{K'}
(\xi _{out}^* )^{S'} [J_{S'K'A'}+(J_1)_{S'K'A'}] |^2 .\label{ce}
\end{eqnarray}

First, we consider the contribution of $J_{S'K'A'}$. Contracting $J_{S'K'A'}$ with $(\xi _{out}^* )^{A'} (\xi _{out}^* )^{K'}(\xi _{out}^* )^{S'}$ gives the following factor
\fa
&& \xi_{(\text{out})}^{*K'} \sigma_{EK'}^{(l)} k_{\text{(tran)}l}  \xi_{(\text{in})}^E = \xi_{(\text{out})}^{*K'}\sigma_{\downarrow K'}^{(l)} k_{\text{(tran)}l}  \xi_{(\text{in})}^\downarrow %
= \xi_{(\text{out})}^{*\uparrow '} \sigma_{\downarrow \uparrow '}^{(l)} k_{\text{(tran)}l} +\xi_{(\text{out})}^{*\downarrow '} \sigma_{\downarrow \downarrow '}^{(l)} k_{\text{(tran)}l}
\nonumber \\
&& ~~~~ =-\sin(\theta/2)\sin\theta-\cos(\theta/2)(\cos\theta-1)=0~,\label{xi3}
\ffa
thus $J_{S'K'A'}$ has no contribution to scattering of the massless wave with spin $\frac{3}{2}$.

Second, we consider the contribution of $(J_1)_{S'K'A'}$.
Taking the contraction of $(J_1)_{S'K'A'}$ and $\xi_{(\text{out})}^{*K'} \xi_{(\text{out})}^{*S'} \xi_{(\text{out})}^{*A'}$,
where * is a functor from spinor category Rep.(1/2,0) of SO(3,1)) to spinor category(Rep.(0,1/2)), and also a map from one spinor space to the corresponding spinor space.
For any spinor $\chi$, it can be expanded in a basis $\{E_A\}$, which is belong to a spinor space of representation(1/2,0), as $\chi=\chi^AE_A$, and the component $\chi^A$ can be obtained via the inner production of $\chi$ and the dual basis $E^A$. 
The complex conjugate of $\chi$ can be formulated as $\chi^*=\sum_{A=\uparrow,\downarrow}(\chi^A)^*E_{A'}$,
where $E_{A'}$ is basis of (0,1/2)-representation of the group SO(3,1), and $(\chi^A)^*$ refers to the complex conjugate of $\chi^A$. Thus, we have $(\chi^A)^*=(\chi^*)^{(A')}$.
% $<E^A,E_B>=\delta^A_B$. 
%The components can be obtained through this operation$\chi^A=<E^A,\chi>$. 

Spinor bases $\xi_{(\text{out})}^E$ are represented by
$\xi_{(\text{out})}^\uparrow\!=\!-\sin(\theta/2)$ and $\xi_{(\text{out})}^\downarrow\!=\!\cos(\theta/2)$. After tedious calculations, we obtain 
\fa
&&\xi_{(\text{out})}^{*K'} \xi_{(\text{out})}^{*S'} \xi_{(\text{out})}^{*A'} (J_{1} )_{S'K'A'}=-\frac{\sqrt2 iM \cos^3 (\theta/2)}{96\pi^2 \omega\sin^2(\theta/2)}~,\label{J1c}
\ffa
here we have made use of Eqs.\,(\ref{m_0}) and (\ref{m_i}).

Collecting Eq.\,\ref(\label{xi3}) and \ref(\label{J1c}), we have
\fa
&& \lim_{M\omega\rightarrow 0} \frac{d\sigma^{(\psi_{\text{sim}(\gamma)}^B)}}{d\Omega} =M^2\frac{\cos^6 (\theta/2)}{\sin^4 (\theta/2)} ~.
\ffa

\subsubsection{Demonstration of equality of simulating wave function and true one}
%is denoted as $dP_{sim}/d\Omega$.
Since \eqref{sim} does not have $\int _{-T/2}^{T/2}dt_0 \exp\{\text{i}(\omega_{sim}-\omega)t_0\}$, i.e., there is no coherent superposition, the differential current of the particles (DCP) in $\Psi_{\text{sim}(\gamma)}^B$ is to small to be considered.

Similarly, the in-going DCP in $\Psi_{\text{sim}(\gamma)}^B$ is too small to be considered. According to Eq.\,(\ref{psi1cB}), the in-going DCP for the sum of $\psi_{\text{sim}(\gamma)}^B$, $\Psi_{\text{sim}(\gamma)}^B$ and $-\psi_{\{1\}(\gamma)}^{(s=\frac{1}{2})B}$ is 0, therefore the in-going DCP in $\psi_{\{1\}(\gamma)}^{(s=\frac{1}{2})B}$ is too small to be considered.

For the out-going DCP in $\psi_{\{1\}(\gamma)}^{(s=\frac{1}{2})B}$, according to Eqs. \eqref{psiEoB}, \eqref{sigmapsi} and \eqref{sigmaPsi}, we have
%We analyze the equation satisfied by $\psi_{\{1\}(\gamma)}^{(s=\frac{1}{2})B}$
%%

\fa
\sigma _{BA'[(\tau) }^{~~} \partial _\beta \psi_{\{1\}(\gamma)]}^{(s=\frac{1}{2})B}
\!+\!\sigma _{BA'[(\beta) }^{~~} \partial _\gamma \psi_{\{1\}(\tau)]}^{(s=\frac{1}{2})B}\!+\!\sigma _{BA'[(\gamma) }^{~~} \partial _\tau \psi_{\{1\}(\beta)]}^{(s=\frac{1}{2})B}\!+\!3(I_{5})_{(\tau)(\beta)(\gamma)A'}=0,\label{wf}
\ffa
with
\begin{eqnarray}
(I_5)_{(\tau)(\beta)(\gamma)A'}=(I_{<3>})_{(\tau)(\beta)(\gamma)A'}\!+\!(I_{<4>})_{(\tau)(\beta)(\gamma)A'}\!
-\!(I_1)_{(\tau)(\beta)(\gamma)A'}\!-\!(Z_1)_{(\tau)(\beta)(\gamma)A'}.\label{ci}
\end{eqnarray}
We can decompose Eq.\,\eqref{wf} into two equations. One is the dynamic equation, corresponding to the index ${(\gamma)\!=\!0}$:
\fa
\sigma _{BA'[(i) }^{~~~~} \partial _j \psi_{\{1\}(0)]}^{(s=\frac{1}{2})B}+\sigma _{BA'[(j) }^{~~~~} \partial _0 \psi_{\{1\}(i)]}^{(s=\frac{1}{2})B}+\sigma _{BA'[(0)}^{~~~~} \partial _i\psi_{\{1\}(j)]}^{(s=\frac{1}{2})B}+3(I_{5})_{(i)(j)(0)A'}=0,\label{psievo}
\ffa
and the other is the constraint equation, corresponding to the index ${(\gamma)\!=\!(1),(2),(3)}$:
\fa
\sigma _{BA'[(i) }^{~~~~} \partial _j \psi_{\{1\}(k)]}^{(s=\frac{1}{2})B}+\sigma _{BA'[(j) }^{~~~~} \partial _k \psi_{\{1\}(i)]}^{(s=\frac{1}{2})B}+\sigma _{BA'[(k) }^{~~~~} \partial _i \psi_{\{1\}(j)]}^{(s=\frac{1}{2})B}+3(I_{5})_{(i)(j)(k)A'}=0.
\ffa

Let $\Lambda[(I_5)_{(\tau)(\beta)(\gamma)A'}] $ denote the set of configurations that satisfy this constraint equation.

Define a functional
\begin{eqnarray}
E[\psi_{(\tau)}^{B}]=\int d^3 x |\psi_{(\tau)}^{B}|^2~.
\end{eqnarray}
After giving the boundary conditions, we can calculate the minimal value of $E[\psi_{(\tau)}^{B}]$ in
 $\Lambda[(I_5)_{(\tau)(\beta)(\gamma)A'}]$, where $\psi_{(\tau)}^{B}$ is denoted as  $\psi_{min(\tau)}^{B}$.
Since $(I_{5})_{(i)(j)(k)A'}$ has only spin-(1/2) component,
when the spatial range is large enough, $\psi_{min(\tau)}^{B}$ does not have spin of $\pm (3/2)$ component. Therefore we can construct an initial state without gravitino.

In order to calculate the out-going DCP in $\psi_{\{1\}(\gamma)}^{(s=\frac{1}{2})B}$, we can write
$\psi_{\{1\}(\gamma)}^{(s=\frac{1}{2})B}$ as the sum of two wave functions
\fa
\psi_{\{1\}(\gamma)}^{(s=\frac{1}{2})B}=\psi_{(I)(\gamma)}^B+\psi_{(II)(\gamma)}^B~.
\ffa
These two functions satisfy the following evolution equations
%$\psi_{(I)(\gamma)}^B$ is
\fa
&&\hskip -1cm \sigma _{BA'[(\tau) }^{~} \partial _\beta \psi_{(I)(\gamma)]}^B+\sigma _{BA'[(\beta) }^{~} \partial _\gamma \psi_{(I)(\tau)]}^B+\sigma _{BA'[(\gamma) } ^{~} \partial _\tau \psi_{(I)(\beta)]}^B+3(I_{5})_{(\tau)(\beta)(\gamma)A'}=0~,\label{SHwf}\\
&&\hskip -1cm \sigma _{BA'[(\tau) }^{~} \partial _\beta \psi_{(II)(\gamma)]}^B+\sigma _{BA'[(\beta) }^{~} \partial _\gamma \psi_{(II)(\tau)]}^B+\sigma _{BA'[(\gamma) }^{~} \partial _\tau \psi_{(II)(\beta)]}^B=0~,\label{plwf}
\ffa
with the initial conditions
\fa
&& \psi_{(I)(\gamma)}^B(-T/2)=\psi_{min(\tau)}^{B}~,\label{df}\\
&& \psi_{(II)(\gamma)}^B(-T/2)=\Psi_{\text{sim}(\gamma)}^B(-T/2)-\psi_{\{1\}(\gamma)}^B(-T/2)-\psi_{min(\gamma)}^B~.\label{plcs}
\ffa
%Since \eqref{wf} and \eqref{SHwf} are consistent, \eqref{plwf} is consistent.

Now we demonstrate that the out-going DCP of $\psi_{(I)(\gamma)}^B$ and $\psi_{(II)(\gamma)}^B$ are either zero or too small to be considered in the domain $\Theta$.

First, let $P_{\Theta}$ be a project operator, which filters out Fourier component outside of $\Theta$. If we use $P_{\Theta}(I_{5})_{(\tau)(\beta)(\gamma)A'}$ to replace $(I_{5})_{(\tau)(\beta)(\gamma)A'}$ in $\eqref{SHwf}$ , then the number of particles received in the domain $\Theta$ remains unchanged.
\begin{eqnarray}
&&\hskip -1cm P_{\Theta} (I_{5})_{(i)(j)(0)A'}=\sigma _{MD'[(i) } \sigma _{NO'(j)]} \epsilon ^{MN} \epsilon ^{O'S'} \epsilon ^{D'K'}
\nonumber\\
&&  \hskip -1cm ~~~~~~\times \Big[\frac{1}{12} \int_{\Theta} d^3k_{(\text{sim})}
\epsilon ^{CG} \sigma _{B(K'}^{(0)} \sigma _{GS'}^{(0)}
\sigma _{CA')}^{(k)} \big(\iota _{(k)}^B+\text{i}\omega\mu_{(k)}^B\big)\text{e}^{\text{i} k_{(\text{sim})l} x^l}\text{e}^{-\text{i} \omega t}
\nn\\
&&  \hskip -1cm ~~~~~~~~~~
+\,\frac{\text{i}}{12} \int_{\Theta}d^3k_{(\text{sim})} \epsilon^{EG}  k_{(\text{sim})l} \sigma_{B(A' }^{(0)} \sigma _{EK'}^{(l) }\sigma _{GS')}^{(k)} \mu _{(k)}^B\text{e}^{\text{i} k_{(\text{sim})l} x^l}\text{e}^{-\text{i} \omega t}\nonumber\\
&&  \hskip -1cm ~~~~~~~~~~
-2\,\int_{\Theta} d^3k_{(\text{sim})}\big[(J_1)_{S'K'A'}+J_{S'K'A'}\big]\text{e}^{\text{i} k_{(\text{sim})l} x^l}\text{e}^{-\text{i} \omega t}\nonumber\\
&& \hskip -1cm ~~~~~~~~~~ + \int_{\Theta} d^3 k_{(\text{sim})}\big[\epsilon _{K’A'} (B_{17} )_{S'}+\epsilon _{S’A'} (B_{18} )_{K'}\big]\text{e}^{\text{i} k_{(\text{sim})l} x^l}\text{e}^{-\text{i} \omega t}\Big]~,\label{dh}
\end{eqnarray}
where $(B_{17})_{S'}$ and $(B_{18})_{K'}$ are complex spinors with spin $\frac{1}{2}$, thus they do not contribute to the scattering of the massless wave with spin $\frac{3}{2}$.

The initial state of $\psi_{(I)(\beta)}^B$ has only spin of $\pm\frac{1}{2}$ components. According to \eqref{goal}, \eqref{SHwf} and \eqref{plwf}, we obtain that the external sources can only inject spin of $\pm\frac{1}{2}$ components into the gravitino field. Therefore there is no scattering amplitude of spin of $\pm\frac{3}{2}$ in $\Theta$ from $\psi_{(I)(\beta)}^B$.

Next, \eqref{plwf} is a homogeneous equation. The out-going particles of
$\psi_{(II)(\beta)}^B$ comes from its in-going particles. We have analyzed that the in-going DCP of  three terms in $\eqref{plcs}$  are too small to be considered, so does the out-going DCP in $\psi_{(II)(\gamma)}^B$. Therefore, the out-going DCP in $\psi_{\{1\}(\gamma)}^{(s=\frac{1}{2})B}$  is too small to be considered.

Summing up the above three sub-steps, we arrive at
\begin{eqnarray}
 \lim_{M\omega\rightarrow 0} \frac{d\sigma}{d\Omega} = M^2\frac{\cos^6 (\theta/2)}{\sin^4 (\theta/2)} ~.
\end{eqnarray}

Notice that this work involves a substantial number of variables and equations, and a summary of the key ones is provided in Appendix.

\section{Conclusion}

%The scattering cross sections of the massless waves with spin $s\!=\!0,\frac{1}{2},1,2$ induced by the spherically-symmetric gravitational field have been extensively studied and the results for the long wavelength limit have been achieved in the literature.
%In this work,
%We employ the simulating wave function method to study the gravitino wave in the spherically-symmetric gravitational field and calculate the scattering cross section in the long wavelength limit. It is proven that the scattering cross section of the long gravitino wave obeys the same pattern as those of the scalar, neutrino, electromagnetic waves. 

We employ the simulating wave function method to analyze the behavior of the gravitino -- a spin-3/2 particle -- in a Schwarzschild field. By deriving the scattering of gravitino wave in the long-wavelength limit, we found that its cross section follows a universal spin-dependent pattern, as other quantum waves do. This work exhibits a underlying symmetry in how different spin quantum waves interact with curved spacetime when their wavelength is much larger than the characteristic size of the gravitational source, and makes a quantitative conclusion for the long-wavelength scattering problem of supersymmetric particles in the spherically symmetric gravitational field.  

\section*{Acknowledgements}
%This work was supported in part by the National Natural Science Foundation of China (Grant Nos. 12475057 and 11973025).
The authors thank Hiroaki Nakajima for helpful discussions and suggestions. This work was supported in part by the National Natural Science Foundation of China (Grant No. 12475057 and 11973025).

%\newpage

\newpage

%\Appendix

\section*{Appendix: Key variables and their relations}
The key variables and related equations are summarized in Table \ref{Table1}.
\begin{table}[!h]
\centering
\begin{tabular}{|m{2.4cm}|m{7.4cm}|m{4cm}|}
\hline
Variable & Meaning  & In equation  \\
\hline
$\psi_{\{1\}(\beta)}^{A}$    &    Perturbation of gravitino wave function      &  \eqref{psi}, \eqref{psiEoB}, \eqref{psi1cB}\\
\hline
$(Z_1)_{(\tau)(\beta)(\gamma)A'}$     & Perturbation induced by
Ricci rotation coefficient and spin connection   &    \eqref{psiEoB}, \eqref{I40}, \eqref{(Z_1)_{(i)(j)(0)A'}}, \eqref{(}, \eqref{psievo}    \\
\hline
$(I_1)_{(\tau)(\beta)(\gamma)A'}$ &Perturbation induced by tetrad basis &\eqref{psiEoB}, \eqref{T8}, \eqref{hskip -}, \eqref{hski}, \eqref{-}, \eqref{psievo}\\

\hline
$\tau_{ijA'}$&Tensor extracted from $(Z_1)_{(\tau)(\beta)(\gamma)A'}$    &\eqref{a}, \eqref{(}    \\
\hline
$J_{S'K'A'}$       & Tensor extracted from $\tau_{ijA'}$   &\eqref{a}, \eqref{J}, \eqref{2},  \eqref{xi3}\\
\hline
$(J_1 )_{S'K'A'}$ &Tensor extracted from $(I_1)_{(\tau)(\beta)(\gamma)A'}$ &\eqref{-}, \eqref{-0}, \eqref{s}, \eqref{psi1cB},  \eqref{xi3}\\
\hline
$\psi_{{scat}(\tau)}^B$  &Scattering wave function in a single direction &     \eqref{psis}, \eqref{psisa}, \eqref{ab}   \\
\hline
          $(I_{<1>})_{(\tau)(\beta)(\gamma)A'}$    &Inhomogeneous term in the field equation for $\psi_{{scat}(\tau)}^B$  &   \eqref{psisa}, \eqref{I_}\\

        \hline
$\Psi_{(\tau)}^B$   &Bounded wave function in a single direction   &\eqref{Psi}, \eqref{I49}, \eqref{sim}      \\
\hline
$\iota_{(\tau)}^B$&Parameter in $\psi_{{scat}(\tau)}^B$ &    \eqref{psis}, \eqref{iota}, \eqref{I_}, \eqref{ae}, \eqref{sigio}, \eqref{ag}, \eqref{goal},  \eqref{cz}, \eqref{cb}, \eqref{cc}, \eqref{dh} \\
\hline
          $(I_{<2>})_{(\tau)(\beta)(\gamma)A'}$ &Inhomogeneous term in the field equation for $\Psi_{(\tau)}^B$ &\eqref{I49}, \eqref{I<2>}, \eqref{I<4>}   \\
\hline
$\psi _{{sim}(\tau)}^B$     & Simulated scattering wave function in all direction   &     \eqref{ab}, \eqref{sigmapsi}, \eqref{sim}, \eqref{bd}    \\
\hline
$(I_{<3>})_{(\tau)(\beta)(\gamma)A'}$ &Inhomogeneous term in the field equation for $\psi _{{sim}(\tau)}^B$ &\eqref{sigmapsi}, \eqref{ad}, \eqref{ae}, \eqref{ag}, , \eqref{ci}\\
\hline
$(I_{<4>})_{(i)(j)(0)A'}$ &Inhomogeneous term in the field equation for $\Psi _{{sim}(\tau)}^B$ & \eqref{I<4>}, \eqref{bz},  \eqref{ci}   \\
\hline

$ \sigma^{(\psi_{{sim}(\gamma)}^B)}$     &Cross section induced by $\psi_{{sim}(\gamma)}^B$    & \eqref{cb}, \eqref{ce}   \\
        \hline
          $(I_5)_{(\tau)(\beta)(\gamma)A'}$ &Inhomogeneous term in the field equation for $\psi_{\{1\}(\tau)}^{(s=\frac{1}{2})B}$ &\eqref{ci}, \eqref{psievo}\\
        \hline

    \end{tabular}
\caption{Key variables, their meaning and related equations}\label{Table1}
\end{table}

The relations for some key variables are described by Figure \ref{figure1}.
\begin{figure}[!h]
\begin{center}
  \includegraphics[width=0.6\linewidth]{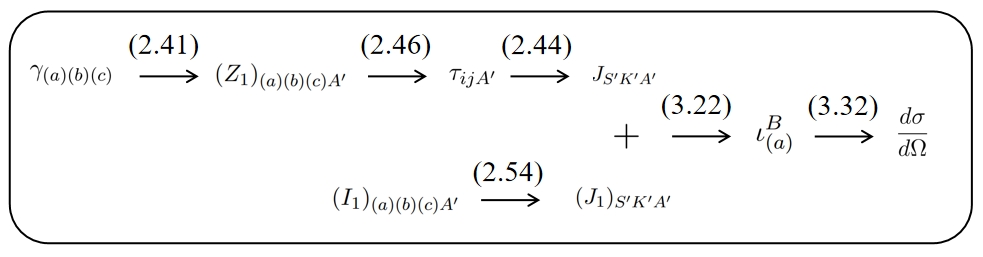}
	\caption{Flow chart for calculating differential scattering cross section of gravitino wave in the long-wavelength limit.}
	\label{figure1}
\end{center}
\end{figure}


%
%\bibitem{b}
%Author, \emph{Title},
%arxiv:1234.5678.
%
%\bibitem{c}
%Author, \emph{Title},
%Publisher (year).

% Please avoid comments such as "For a review'', "For some examples",
% "and references therein" or move them in the text. In general,
% please leave only references in the bibliography and move all
% accessory text in footnotes.

% Also, please have only one work for each \bibitem.

%\end{thebibliography}

\end{document}